\documentclass[epj]{svjour}
\usepackage{latexsym}
\usepackage[title]{appendix}
\usepackage{graphicx}
\usepackage{float}
\usepackage{amsmath}
\usepackage{hyperref}
\usepackage[caption=false]{subfig}
\usepackage{bm}
\usepackage[numbers,sort&compress]{natbib}

\begin{document}

\title{Including Millisecond Pulsars inside the Core of Globular Clusters in Pulsar Timing Arrays}

\author{Michele Maiorano\inst{1,2}\thanks{ORCID: 0000-0003-3066-3369}\thanks{Correspondence: michele.maiorano@le.infn.it} \and Francesco de Paolis\inst{1,2,3} \and Achille Nucita\inst{1,2,3}}

\institute{Department of Mathematics and Physics “Ennio De Giorgi”, University of Salento, Via per Arnesano, I-73100 Lecce, Italy \and Istituto Nazionale di Fisica Nucleare INFN, Sezione di Lecce, Via per Arnesano, I-73100 Lecce, Italy \and Istituto Nazionale di Astrofisica INAF, Sezione di Lecce, Via per Arnesano, I-73100 Lecce, Italy}

\date{Received: date / Revised version: date}

\abstract{We suggest the possibility of including millisecond pulsars inside the core of globular clusters in pulsar timing array experiments. Since they are very close to each other, their gravitational wave induced timing residuals are expected to be almost the same, because both the Earth and the pulsar terms are correlated. We simulate the expected timing residuals, due to the gravitational wave signal emitted by a uniform supermassive black-hole binary population, on the millisecond pulsars inside a globular cluster core. In this respect, Terzan 5 has been adopted as a globular cluster prototype and, in our simulations, we adopted similar distance, core radius, and number of millisecond pulsars contained in it. Our results show that the presence of a strong correlation between the timing residuals of the globular cluster core millisecond pulsars can provide a remarkable gravitational wave signature. This result can be therefore exploited for the detection of gravitational waves through pulsar timing, especially in conjunction with the standard cross-correlation search carried out by the pulsar timing array collaborations.
\PACS{
      {PACS-key}{discribing text of that key}   \and
      {PACS-key}{discribing text of that key}
     }    
}
\maketitle

\section{Introduction}
\label{intro}
The existence of gravitational waves (GWs) has been predicted for the first time by Albert Einstein's theory of General Relativity \cite{einstein1916}. Within this context, the geodesic deviation equation provides the tools to detect GWs, because it implies that GWs strain the space by a factor proportional to the GW amplitude $h$. Originally, due to the smallness of this effect, the scientific community was in agreement on the fact that the GW detection was nearly impossible. Despite that, in the years, many detection techniques have been developed, and eventually, the first detection of GWs (\emph{i.e.} the GW150914 event) emitted during the coalescence of two black-holes (BHs), with inferred masses of $36$ M$_{\odot}$ and $29$ M$_{\odot}$, was achieved in 2015 \cite{abbott2016} by the LIGO (Laser Interferometer Gravitational-Wave Observatory) collaboration, using the Livingston and Hanford gravitational interferometers. This opened up a new window for observational astronomy and the detection of the first binary neutron star merging event (\emph{i.e.} the GW170817 event) \cite{abbott2017} marked the beginning of the multi-messenger astronomy era.

Ground-based laser interferometers are sensitive only to high-frequency (\emph{i.e.} in the frequency range $\simeq 10\text{--}10^3$ Hz) GWs and, for this reason, the best detectable GW sources are black-hole binaries (BHBs) or neutron star binaries (NSBs) at the final stage of their evolution \cite{abbott2017}. To observe these objects in earlier phases, along with other types of low-frequency (\emph{i.e.} in the frequency range $\simeq 10^{-5}\text{--}1$ Hz) GW sources, like extreme mass-ratio inspirals (EMRIs) and massive binaries (MBs), space-based laser interferometers (that bypass the problems due to the Earth seismic noise) are needed. The most promising project in this direction is LISA (Laser Interferometer Space Antenna), a constellation of three satellites, separated by about $1.5\times 10^6$ km, in a triangular configuration, orbiting with Earth around the Sun \cite{lisa2017}.

Although LISA, whose launch is planned in the early 2034 (see ref. \cite{lisaweb}), will be an extremely advanced detector, its sensitivity will not probably allow the detection of ultra--low-frequency (\emph{i.e.} in the frequency range $\simeq 10^{-10}\text{--}10^{-6}$ Hz) GWs. These GWs can be generated by many sources of cosmological interest, like supermassive black-hole binaries (SMBHBs) \cite{rajagopal1995} or cosmic strings \cite{damour2001}. The opportunity to detect such GWs is offered by Pulsar Timing Arrays (PTAs), which are experiments that consist in constant monitoring the radio emission from the most stable isolated and binary millisecond pulsars (MSPs) across the sky with the aim to detect variations commonly known as timing residuals, possibly induced by  GWs \cite{sazhin1978,detweiler1979}, between the observed times of arrival (ToAs) of each pulse and those expected by the timing model.

PTAs should already have the potential to detect the stochastic gravitational wave background (GWB) due to the ensemble of GWs emitted by the SMBHB population and, in fact, an interesting common-spectrum process compatible with it has been recently found \cite{arzoumanian2020, goncharov2021}.  However, the true nature of this signal is unclear, so other interpretations (see, \emph{e.g.}, ref. \cite{blasi2021,ratzinger2021,liang2021}) are still viable. For this reason, it has not been possible to claim the GWB detection yet.

The GW-induced timing residuals depend on the difference between the metric perturbation at the position of the observer (\emph{i.e.} the Earth term) and the metric perturbation at the position of the pulsar (\emph{i.e.} the pulsar term). This means that even if all the MSPs lie in the same GWB, in general, their GW-induced timing residuals are different. Therefore, the current GWB search is based on finding the quadrupolar cross-correlation in the timing residuals described by the Hellings \& Downs function \cite{hellings1983}. Since the pulsar term gives an uncorrelated contribution to timing residuals, it is usually treated as an additional source of white noise. 

The considerations above imply also that the GW-induced timing residuals for MSPs located at almost the same position are expected to be approximately the same. Such effect can only be observed inside a globular cluster (GCs) core where tens of MSPs are confined, in most cases, within a very small distance from its center.

Here, we discuss some good reasons for including GC core MSPs in PTAs. First of all, the interest in these MSPs is well motivated by the fact that the GWB contribution in their timing residuals should be very similar for each of them and, therefore, observing a strong correlation would give an additional “smoking gun” for the GWB detection. Moreover, such a correlation can be used to discriminate the GW-induced timing residuals, either in the case of the GWB or in the case of the continuous GWs, from other effects that produce timing residuals of the same order of magnitude but which turn out to be different for each MSP\footnote{In some rare cases an MSP can also have a planetary companion (or even more than one) which may be responsible for low-frequency timing residuals. The first exoplanet was indeed discovered just thanks to that effect \cite{wolszczan1992}.}\cite{phinney1992,phinney1993}. Eventually, having a set of MSPs so close to each other can be helpful in the standard quadrupolar cross-correlation search because it can give useful information on the small-angle region of the Hellings \& Downs function\footnote{GC MSPs might also be important for the gravitational bursts with memory (BWMs) detection. Indeed, since GCs are often characterized by a high stellar density, this makes them suitable for the occurrence of BWM. Such events would induce impact parameter-dependent timing residuals on all GC MSPs (see ref. \cite{madison2017} for a more detailed discussion)}.

Even though the advantages of including GC core MSPs in PTAs appear very appealing, we must point out that the timing of GC MSPs over a long time spans is much more complicated than for Galactic MSPs. Timing perturbations from accelerations due to the overall GC gravitational potential, as well as nearby stellar motions, tend to dominate the low-frequency timing properties of the GC MSPs \cite{phinney1992,phinney1993}. Therefore, GC MSP usually have non-negligible second-order spin-period time derivative (see, \emph{e.g.}, ref. \cite{freire2017}) and, in some cases, even higher-order time derivative \cite{prager2017}. This implies that, without accurate modeling of all these effects, GC MSPs are not adequate for GW detection. At the present state of the art, we still do not have a very efficient way to deal with GC MSPs, but more accurate timing models might be built in the future thanks to more precise multi-wavelength observations as well as new theoretical developments.

\section{Theoretical Aspects}
\label{sec:1}
\subsection{Reference frame}
\label{subsec:1}
Let us first start by fixing the origin of our $Oxyz$ reference frame in the Solar System barycenter\footnote{Note that the MSP radio emission is actually observed from the Earth, which is not an inertial reference frame. Therefore, the ToA data are transformed to the SSB reference frame.} (SSB). Within the PTA context, the SSB reference frame can be considered, in good approximation, an inertial reference system, and its coordinates are known with great precision. In this reference frame, we indicate the versor in the direction of an MSP as:
\begin{equation}
p^i=(\sin\theta_p\cos\phi_p,\sin\theta_p\sin\phi_p,\cos\theta_p)
\label{mspposition}
\end{equation}
where $\theta_p$ is the angle, in the $yz$ plane, between the $z$ axis and $p^i$ and $\phi_p$ is the angle, in the $xy$ plane, between the $x$ axis and $p^i$. We also indicate the versor in the direction of the GWs emitted by the SMBHB with:
\begin{equation}
\Omega^i=(\sin\theta\cos\phi,\sin\theta\sin\phi,\cos\theta)
\label{gwsposition}
\end{equation}
where $\theta$ is the angle, in the $yz$ plane, between the $z$ axis and $\Omega^i$ and $\phi$ is the angle, in the $xy$ plane, between the $x$ axis and $\Omega^i$. Since we are considering SMBHB sources at distance much higher than that of the MSPs, the GW wave-front can be considered, in good approximation, as a plane wave and it can be fully identified by the orthogonal versors
\begin{equation}
\begin{split}
m^i &=(\sin\phi,-\cos\phi,0)\\
n^i &=(\cos\theta\cos\phi,\cos\theta\sin\phi,-\sin\theta)
\end{split}
\label{mnversors}
\end{equation}

\subsection{Timing residuals induced by gravitational waves}
\label{subsec:2}
A GW can stretch and compress the space, causing a delay or an advance in the pulse ToAs. Therefore it can lead to a variation in the observed frequency of the MSP, which is often referred as pulse redshift, given by\footnote{For an exhaustive derivation see ref. \cite{maggiore2008b}.}:
\begin{equation}
z(t,\Omega^i)=\sum_{A=+,\times}F^A[h^A(t,x^i=0)-h^A(t-t_p,x^i=t_p p^i)]
\label{pulseredshift}
\end{equation}
where $t_p$ is the pulse time of flight\footnote{In this paper geometrical units c=G=1 have been adopted.} from the MSP position $x_p^i$ to the SSB, $h^A$ is the $A$-th GW polarization state amplitude and $F^A$ is the antenna pattern function, given by:
\begin{equation}
F^A=\frac{1}{2}\frac{p^ip^je_{ij}^A}{(1+\Omega_ip^i)}
\label{antennapatternfunction}
\end{equation}
where $e_{ij}^A$ is the GW polarization tensor of the $A$-th GW polarization state\footnote{In this paper the Einstein notation, which indicates the sum over repeated indices, has been adopted.}. In the most general case, that is when more than one GW has to be considered, the GW polarization tensor can be expressed as:
\begin{equation}
\begin{split}
e_{ij}^+ &= m_im_j-n_in_j\\
e_{ij}^\times &= m_in_j+n_im_j
\end{split}
\label{polarizationtensor}
\end{equation}
The GW polarization tensor components have been determined in appx. \ref{app:1}. In eq. \eqref{pulseredshift} the two terms in the square brackets are often called, respectively, the Earth and pulsar terms since the former indicates the metric perturbation in the proximity of the Earth while the latter refers to the MSP. By integrating eq. \eqref{pulseredshift} with respect to time, one can define the function
\begin{equation}
r(t,\Omega^i)=\int_0^tdt \ 'z(t',\Omega^i)
\label{gravitationaltimingresidual}
\end{equation}
which describes the GW-induced timing residuals.

\subsection{Supermassive black-hole binary gravitational waves}
\label{subsec:3}
The GW emission from an SMBHB starts to become effective and possibly detectable when the separation distance between the two merging BHs reaches sub-parsec scales. In the case of BHBs in circular orbits, the GW emission is characterized by an amplitude $h$, given by
\begin{equation}
h=\sqrt{\frac{32}{5}}\frac{\mathcal{M}^{5/3}}{D}\left(\frac{2\pi f}{1+z}\right)^{2/3}
\label{smbhbgwamplitude}
\end{equation}
where $f$ is the GW frequency, $z$ is the cosmological redshift of the SMBHB, $D$ is its luminosity distance from the SSB and $\mathcal{M}$ is the so-called chirp mass. Using eq. \eqref{smbhbgwamplitude}, $h^A$ can be expressed as
\begin{equation}
h^A=\operatorname{Re}\lbrace he^{j(\omega t+k_ix^i+\alpha^A)}\rbrace
\label{ha}
\end{equation}
where $j=\sqrt{-1}$ is the imaginary unit, $\operatorname{Re}$ denotes the real part of the expression in brackets, $\omega=2\pi f$ is the GW angular frequency, $k^i=\omega\Omega^i$ is the GW wavenumber and $\alpha^A$ is the initial phase of the $A$-th GW polarization state. Using eqs. \eqref{pulseredshift} and \eqref{ha}, the Earth term can be expressed as:
\begin{equation}
h^A(t,x^i=0)=\operatorname{Re}\lbrace he^{j(\omega t+\alpha^A)} \rbrace
\label{earthterm}
\end{equation}
and the pulsar term as:
\begin{equation}
h^A(t-t_p,x^i=t_p p^i)=\operatorname{Re}\left\lbrace he^{j(\omega t -\omega t_p(1-\Omega_i p^i)+\alpha^A)}\right\rbrace
\label{pulsarterm}
\end{equation}

In general, the GW frequency of the Earth term in eq. \eqref{earthterm} and that of the pulsar term in eq. \eqref{pulsarterm} are slightly different because, during the pulse time of flight, a GW frequency evolution occurs, due to the SMBHB shrinking caused by the energy loss by GW emission. It follows that the GW frequency evolution is negligible when it takes an amount of time much larger than the pulse time of flight, namely:
\begin{equation}
t_p\ll\frac{5}{256}\left(\frac{\mathcal{M}G}{c^3}\right)^{-5/3}\left[\omega^{-8/3}\right]_{\Delta\omega} 
\label{frequencyevolution}
\end{equation}
where the right member indicates the time interval required to change the GW frequency by an amount $\Delta\omega$ \cite{shapiro1983}. Therefore, eq. \eqref{frequencyevolution} implies that most SMBHBs emitting ultra--low-frequency GWs, which represent some of the main targets of PTAs, are not expected to show a significant GW frequency evolution.

Using eqs. \eqref{earthterm} and \eqref{pulsarterm}, the GW-induced timing residuals in eq. \eqref{gravitationaltimingresidual} can be expressed as 
\begin{equation}
r(t,\Omega^i)=\operatorname{Re}\left\lbrace\frac{h}{j\omega}\sum_{A=+,\times}F^A\left[e^{j(\omega t-\alpha^A)}-e^{j(\omega t -\omega t_p(1-\Omega_i p^i)-\alpha^A)}\right]_0^t\right\rbrace
\label{gravitationaltimingresidualintegrated}
\end{equation}

\section{Simulated timing residuals}
\label{sec:2}
Currently, the main PTA collaborations are the European Pulsar Timing Array (EPTA) \cite{desvignes2016}, the Indian Pulsar Timing Array (InPTA) \cite{joshi2018}, the North American Nanohertz Observatory for Gravitational Waves (NANOGrav) \cite{arzoumanian2018}, and the Parkes Pulsar Timing Array (PPTA) \cite{reardon2016}, and all of them are part of the International Pulsar Timing Array (IPTA) \cite{verbiest2016}. The PTA MSPs are just over 10\% of the known MSPs \cite{perera2019}, which are more than $400$ \cite{atnfweb,manchester2005}. Unfortunately, the majority of MSPs are considered, at present, not suitable for GW searches because of their timing properties. In particular, among all the known GC MSPs \cite{freireweb}, only $B1821-24A$, also known as $J1824-2452A$, which is an energetic pulsar visible in radio, X-rays and $\gamma$-rays (see, \emph{e.g.}, ref. \cite{lyne1987,saito1997,wu2013,bilous2015}) lying in the M28 GC core, is currently included in PTAs. However, according to PPTA observations, this MSP is characterized by one of the largest timing and dispersion measure variation noise levels of any MSPs, and is observed only with low priority\footnote{Interestingly enough, $B1821-24A$ is also characterized by a large value of the second-order spin-frequency derivative (\emph{i.e.} $\ddot{\nu}=29.42\pm 15.75\times 10^{-27}\,{\rm s}^{-3}$) but seems to be affected only slightly, probably at a level not much larger than $15\%$, by the GC gravitational potential \cite{liu2019}.} \cite{kerr2020}.

\begin{figure}[H]
\centering
\includegraphics[scale=0.65]{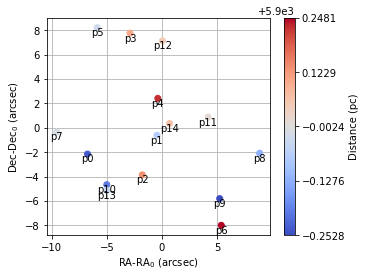}\hfill
\caption{MSP set simulation. A simulated set of 15 MSPs, uniformly distributed inside the core of a GC located at a distance of $5.9$ kpc and characterized by a core angular radius of $0.16$ arcmin is shown. The dots indicate the MSP coordinates (RA and Dec) with respect to GC center (RA$_0$ and Dec$_0$), expressed in arcsec, while the dot colors indicate, as described by the color bar, the MSP distance with respect to GC center, expressed in pc. The MSP distance with respect to SSB can be obtained by summing the values indicated by the labels of the color bar with the value on its top.}
\label{fig:gcmsp}
\end{figure}

\begin{figure}[H]
\centering
\includegraphics[scale=0.65]{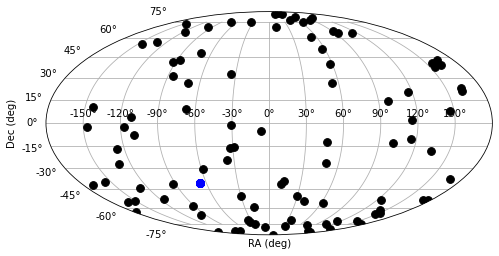}\hfill
\caption{SMBHB set simulation. A simulated set of $100$ SMBHBs, uniformly distributed all over the sky, is shown. The black dots indicate the position of the SMBHBs while the blue dot indicates the position of the considered GC with the 15 simulated MSPs within its core radius. Galactic coordinates RA and Dec, in degrees, are adopted in the plot.}
\label{fig:skymap}
\end{figure}

\begin{figure}[H]
\centering
\subfloat[SMBHB chirp mass distribution. \label{a}]{\includegraphics[scale=0.55]{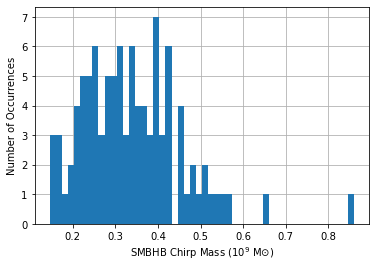}}\hfill
\subfloat[SMBHB distance distribution. \label{b}]{\includegraphics[scale=0.55]{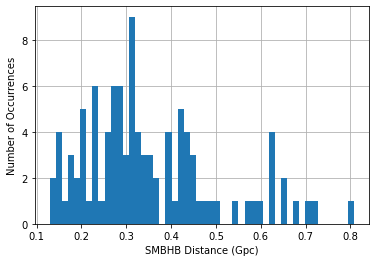}}

\subfloat[GW frequency distribution. \label{c}]{\includegraphics[scale=0.55]{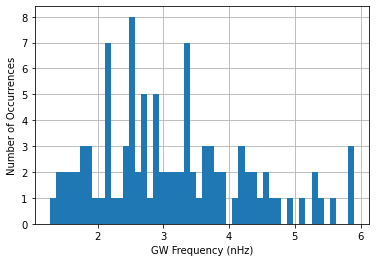}}\hfill
\subfloat[GW strain distribution. \label{d}]{\includegraphics[scale=0.55]{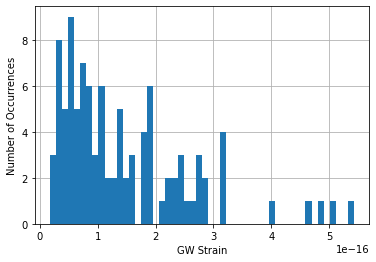}}
\caption{SMBHB parameter simulation. In panel \protect\subref{a} the SMBHB chirp mass (in units of $10^9\,M_{\odot}$) log-normal distribution is shown. In panel \protect\subref{b} the SMBHB distance (in Gpc) log-normal distribution is shown. In panel \protect\subref{c} the GW frequency (in nHZ units) log-normal distribution is shown. In panel \protect\subref{d} the GW strain distribution (in units of $10^{-16}$) is shown. On the vertical axis of each panel the number of occurrences is plotted, see sec. \ref{sec:2} for the details about the parameter distribution of the GW sources.}

\label{fig:smbhb}
\end{figure}

\begin{figure}[H]
\centering
\subfloat[Simulated $30$ years GW-induced timing residuals. \label{30}]{\includegraphics[scale=0.65]{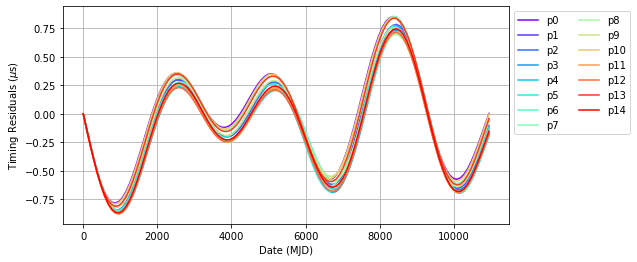}}\hfill

\subfloat[Simulated $10$ years GW-induced timing residuals. \label{10}]{\includegraphics[scale=0.65]{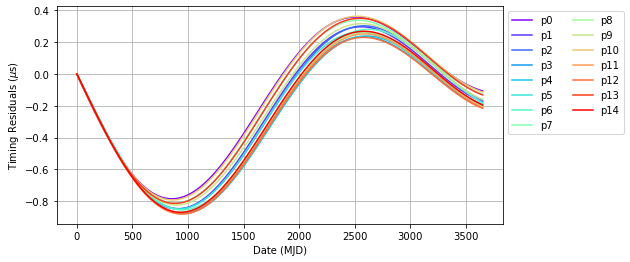}}\hfill
\caption{GW-induced timing residuals. In panel \protect\subref{30} the GW-induced timing residuals due to the overall contribution of the GWs emitted by all the simulated SMBHBs, for each simulated MSP, are plotted over a time interval of $30$ years. In panel \protect\subref{10} the GW-induced timing residuals are plotted over a time interval of only $10$ years. The colored curves indicate the GW-induced timing residuals for each of the 15 simulated MSPs, as described by the legend. On the horizontal axis is plotted the date, expressed in MJD, measured from the arbitrary beginning of the simulated ToA observation. On the vertical axis of both panels are plotted the timing residuals, expressed in $\mu$s.}
\label{fig:timingresiduals}
\end{figure}

\begin{figure}[H]
\centering
\includegraphics[scale=0.65]{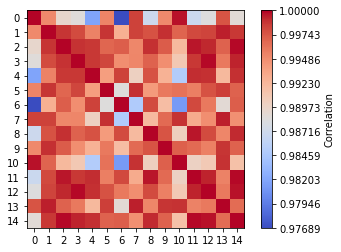}\hfill
\caption{Pearson correlation matrix. The Pearson correlation matrix between the GW-induced timing residuals is shown for each pair of the simulated MSPs. The colored elements of the matrix correspond to each pair of the simulated MSPs, and the colors indicate the values of the correlation, as described by the color bar. On the horizontal and vertical axes, we plotted the natural numbers labeling each MSP. As expected for this set of MSPs, the Pearson correlation value is always close to unity due to the fact that the timing residuals show approximately the same time dependence.}
\label{fig:correlation}
\end{figure}

\begin{figure}[H]
\centering
\includegraphics[scale=0.55]{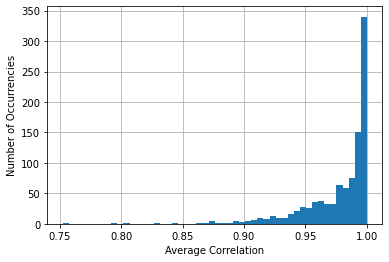}\hfill
\caption{Average, or mean value, of the Pearson correlation coefficient distribution. On the horizontal axis is plotted the average Pearson correlation coefficient, while the number of occurrences is plotted on the vertical axis.}
\label{fig:averagecorrelation}
\end{figure}

In the following part of the paper we present some good reasons why they may be actually very good targets, and any effort should be done to include them in current and future PTAs.
In order to convince the reader about the advantages of considering GC core MSPs, we first simulated a set of $15$ MSPs, inside the core of a GC located at a distance of $5.9$ kpc and characterized by an angular radius of $0.16$ arcmin. The choice of these parameters has been made by adopting the Terzan 5 GC as a prototype because it is the most populated GC \cite{cadelano2018}. We randomly extracted, from a uniform distribution\footnote{In accordance with what is described by the GC core mass-density distribution \cite{binney1987}.}, the position of 15 GC core MSPs (see fig. \ref{fig:gcmsp}).
We also simulated a set of SMBHBs emitting continuous GWs. We randomly extracted, from a uniform distribution, the position of the SMBHBs in the sky (see fig. \ref{fig:skymap}). We also randomly extracted, from a log-normal distribution, the chirp mass of the SMBHBs, their distance, and the GW emission frequency, respectively, in the ranges\footnote{The distribution parameters have been arbitrarily chosen with the aim to simulate the GW emission from a possible SMBHB population observable by PTAs. The results in refs. \cite{tucci2017,celoria2018,sanchis2021} have been used as reference.} $10^8\text{--}10^9$ M$_\odot$, $10^8\text{--}10^9$ pc, and $10^{-9}\text{--}10^{-8}$ Hz \cite{tucci2017,celoria2018,sanchis2021}. We then used eq. \eqref{smbhbgwamplitude} to obtain the GW strain distribution (see fig. \ref{fig:smbhb}).
We finally calculated, using eq. \eqref{gravitationaltimingresidualintegrated}, the GW-induced timing residuals due to the overall contribution of the GWs emitted by all the simulated SMBHBs, for each simulated MSP. We considered both the GW polarization states, fixing $\alpha^+=0$ and $\alpha^\times=\pi/2$. The obtained results are shown in fig. \ref{fig:timingresiduals}.
In order to quantify the correlation of the timing residuals among the 15 simulated MSPs, we calculated the Pearson correlation matrix between the GW-induced timing residuals for each pair of the simulated MSPs \footnote{Note that the Pearson correlation matrix coefficients are in the interval [-1,1], where 1 means perfect correlation, 0 means non-correlation and -1 means perfect anti-correlation.}. The diagonal elements of the Pearson correlation matrix are unitary: indeed, these elements indicate the correlation of each curve with itself (see fig. \ref{fig:correlation}).
We produced $1000$ different simulations and, for each of them, we calculated the mean value of the Pearson correlation coefficients\footnote{The mean value has been calculated by ignoring the diagonal elements of the matrix.} (see fig. \ref{fig:averagecorrelation}).

\section{Conclusions}
\label{sec:3}
In this paper, we simulated the GW-induced timing residuals, due to the overall contribution of the GWs emitted by a set of SMBHBs, for each of the 15 MSPs inside the core of a GC modeled adopting the Terzan 5 GC as a prototype. As one can see from figs. \ref{fig:timingresiduals} and \ref{fig:correlation}, all the GW-induced timing residuals on the 15 MSPs have almost the same shape and, moreover, the Pearson correlation coefficients calculated for each pair are almost of the order of the unity. In all the $1000$ simulations we made, as shown in fig. \ref{fig:averagecorrelation}, the mean value of the Pearson correlation coefficients are in the range $\simeq 0.75\text{--}1$ and, for the large majority of the simulated systems, are even greater than about $0.95$. This means that the GW-induced timing residuals are highly correlated and that the GC core MSPs can be powerful tools for detecting GWs. Indeed, the observation of such correlated timing residuals would be strong evidence of the effect of GWs on the timing of MSPs.

Moreover, using this class of MSPs, and in particular the cross-correlation between their timing residuals, as regular PTA MSPs, can give useful information on the small-angle region of the Hellings \& Downs function.

A further possibility, that will be the subject of a paper in preparation, is to use the GC core MSPs as if it was a single MSP, by averaging the GW-induced timing residuals of all the MSPs. This could provide a helpful way to emphasize the correlation and reduce the contribution of the uncorrelated noise that affects any real observation. Applying this procedure to different GCs it might be possible to perform the standard quadrupolar cross-correlation search. If a GWB signature emerges, the GC core MSPs with the highest mean correlation coefficient can be studied individually via periodic analysis algorithms.

It is important to remark that in the analysis above, we considered an ideal situation in order to better highlight the main advantages of including in the PTAs the GC core MSPs. However, we are aware that the details of the GC gravitational potential and the actual MSP positions with respect to its center, as well as the dynamics of the other GC components, may play an important role (see, \emph{e.g.}, refs. \cite{phinney1992,phinney1993,freire2017,prager2017,depa1996,abbate2019}). Indeed, we are planning to take into account all the induced effects in a later paper. It should also be noticed that, currently, this class of MSPs is not routinely timed with a precision adequate for the detection of GW-induced timing residuals. Nevertheless, the situation might change in the near future, thanks to new powerful detectors, like the Five hundred meter Aperture Spherical Telescope (FAST) \cite{nan2011}, the MeerKAT radio telescope \cite{bailes2020} and the Square Kilometre Array (SKA) \cite{weltman2020}. SKA, in particular, will certainly play a crucial role in ultra--low-frequency GW detection. With its unprecedented sensitivity, it will be possible to significantly improve the timing of MSPs. Eventually, building more accurate timing models for GC MSPs might open, hopefully as soon as possible, the possibility of adopting the strategy proposed in this paper.

\section*{Declarations}
\subsection*{Funding}
No funding was received.
\subsection*{Conflicts of interest/Competing interests}
The authors declare they have no financial interests.
\subsection*{Availability of data and material}
Only simulated data have been used in the paper.
\subsection*{Code availability}
The numerical codes are available on request.
\subsection*{Authors' contributions}
All authors equally contributed to the paper. The first draft of the manuscript was written by Michele Maiorano and all authors commented on previous versions of the manuscript. All authors read and approved the final manuscript.
\subsection*{Additional declarations for articles in life science journals that report the results of studies involving humans and/or animals}
Not applicable.
\subsection*{Ethics approval}
Not applicable.
\subsection*{Consent to participate}
The authors give their consent.
\subsection*{Consent for publication}
The authors give their consent for publication.

\section*{Acknowledgements}
We warmly acknowledge Andrea Possenti, of the Istituto Nazionale di Astrofisica (INAF), for many useful discussions. We also acknowledge the support of the Theoretical Astroparticle Physics (TAsP) and Euclid projects of the Istituto Nazionale di Fisica Nucleare (INFN). We thank the Referee for the useful comments.

\begin{appendices}

\section{The Polarization Tensor Components}
\label{app:1}
The GW polarization tensor components of the $+$ polarization state and the $\times$ polarization state can be determined from eqs. \eqref{polarizationtensor}. Using eqs. \eqref{mspposition} and \eqref{gwsposition} one obtains:
\begin{equation}
\begin{split}
e_{11}^+ &= (\sin^2\phi)-(\cos^2\theta)(\cos^2\phi)\\
e_{22}^+ &= (\cos^2\phi)-(\cos^2\theta)(\sin^2\phi)\\
e_{33}^+ &= -(\sin^2\theta)\\
e_{12}^+ &= -\sin\phi\cos\phi(\cos^2\theta+1)\\
e_{23}^+ &= \sin\theta\cos\theta\cos\phi\\
e_{13}^+ &= \sin\theta\cos\theta\sin\phi\\
e_{11}^\times &= 2\cos\theta\sin\phi\cos\phi\\
e_{22}^\times &= -2\cos\theta\sin\phi\cos\phi\\
e_{33}^\times &= 0\\
e_{12}^\times &= \cos\theta(\sin^2\phi-\cos^2\phi)\\
e_{23}^\times &= \sin\theta\cos\phi\\
e_{13}^\times &= -\sin\theta\sin\phi\\
\end{split}
\label{polarizationtensorplus}
\end{equation}
\end{appendices}

\bibliographystyle{unsrt}
\bibliography{bibliography.bib}
\end{document}